\documentclass[a4paper,fleqn]{cas-dc}

\usepackage[sorting=none]{biblatex}
\usepackage{hyperref}

\makeatletter                                                                   
\newlength{\bibsep}{\@listi \global\bibsep\itemsep \global\advance\bibsep by\parsep}
\makeatother   

\begin{document}

%\let\WriteBookmarks\relax
%\def\floatpagepagefraction{1}
%\def\textpagefraction{.001}

% Short title
%\shorttitle

% Short author
\shortauthors{Rafael Pineda M., William J. Herrera}

% Main title of the paper
\title [mode = title]{Edge states, transport and topological properties of hererostructures in the SSH model}                      
% Title footnote mark
% eg: \tnotemark[1]
%\tnotemark[1,2]

% Title footnote 1.
% eg: \tnotetext[1]{Title footnote text}
% \tnotetext[<tnote number>]{<tnote text>} 
%\tnotetext[1]{This document is the results of the research
  % project funded by the National Science Foundation.}

%\tnotetext[2]{The second title footnote which is a longer text matter
  % to fill through the whole text width and overflow into
  % another line in the footnotes area of the first page.}

% First author
%
% Options: Use if required
% eg: \author[1,3]{Author Name}[type=editor,
%       style=chinese,
%       auid=000,
%       bioid=1,
%       prefix=Sir,
%       orcid=0000-0000-0000-0000,
%       facebook=<facebook id>,
%       twitter=<twitter id>,
%       linkedin=<linkedin id>,
%       gplus=<gplus id>]
\author{Rafael Pineda M}%[type=editor,
                        %auid=000,bioid=1,
                       %prefix=Sir,
                       % role=.,
                        %orcid=0000-0001-7511-2910]
%]
% Corresponding author indication
%\cormark[1]

% Footnote of the first author
%\fnmark[1]

% Email id of the first author
\ead{raapinedame@unal.edu.co}

% URL of the first author
%\ead[url]{www.cvr.cc, cvr@sayahna.org}

%  Credit authorship
%\credit{Conceptualization of this study, Methodology, Software}

% Address/affiliation
%\affiliation[1]{organization={Universidad Nacional de Colombia},
    %addressline={Carrera 45 26-85}, 
    %city={Bogota},
    % citysep={}, % Uncomment if no comma needed between city and postcode
    %postcode={111321}, 
    % state={},
   % country={Colombia}}

% Second author
\author{William J. Herrera}%[style=chinese]

% Third author
%\author[2,3]{CV Rajagopal}[%
 %  role=Co-ordinator,
 %  suffix=Jr,
 %  ]
%\fnmark[2]
\ead{jherreraw@unal.edu.co
}
%\ead[URL]{www.sayahna.org}

%\credit{Data curation, Writing - Original draft preparation}

% Address/affiliation
%\affiliation[2]{organization={Sayahna Foundation},
    % addressline={}, 
  %  city={Jagathy},
    % citysep={}, % Uncomment if no comma needed between city and postcode
   % postcode={695014}, 
%    state={Trivandrum},
 %   country={India}}

% Fourth author
%\author%
%[1,3]
%{Rishi T.}
%\cormark[2]
%\fnmark[1,3]
%\ead{rishi@stmdocs.in}
%\ead[URL]{www.stmdocs.in}

%\affiliation[3]{organization={STM Document Engineering Pvt Ltd.},
%    addressline={Mepukada}, 
%    city={Malayinkil},
    % citysep={}, % Uncomment if no comma needed between city and postcode
%    postcode={695571}, 
%    state={Trivandrum},
%    country={India}}

% Corresponding author text
%\cortext[cor1]{Corresponding author}
%\cortext[cor2]{Principal corresponding author}

% Footnote text
%\fntext[fn1]{This is the first author footnote. but is common to third
 % author as well.}
%\fntext[fn2]{Another author footnote, this is a very long footnote and
 % it should be a really long footnote. But this footnote is not yet
 % sufficiently long enough to make two lines of footnote text.}

% For a title note without a number/mark
%\nonumnote{This note has no numbers. In this work we demonstrate $a_b$
%  the formation Y\_1 of a new type of polariton on the interface
 % between a cuprous oxide slab and a polystyrene micro-sphere placed
 % on the slab.
 %}

% Here goes the abstract
\begin{abstract}
In this paper, we discuss the topological and transport properties of binary heterostructures of different topological materials. Using the Su-Schrieffer-Heeger model and the method of Green's functions, we calculate and characterize the edge states of these heterostructures. We study the topological phase through an invariant calculated from the Zak phase in order to build phase diagrams. We obtained an analytical result that allows us to find the topological phase diagram for a one-dimensional general chain and for heterostructures from the cross band condition. We also calculate the differential conductance with the non-equilibrium Green function technique showing the tunneling of the edge states and discussing its possible design and experimental application. 

%\noindent\texttt{\textbackslash begin{abstract}} \dots 
%\texttt{\textbackslash end{abstract}} and
%\verb+\begin{keyword}+ %\verb+...+ \verb+\end{keyword}+ 
%which
%contain the abstract and keywords respectively. 

\end{abstract}

% Use if graphical abstract is present
% \begin{graphicalabstract}
% \includegraphics{figs/grabs.pdf}
% \end{graphicalabstract}

% Research highlights

% Keywords
% Each keyword is seperated by \sep
%\begin{keywords}
%One dimensional Tight binding model \sep Topological Heterostructures \sep Topological superlattice \sep Edge states \sep Phase Diagrams
%\end{keywords}

\maketitle

%\phantomsection
\section{Introduction}

Progress in condensed matter physics focuses on the synthesis of new materials with special properties, such as topological insulators, which are classified as a new quantum phase of matter \cite{RevModPhys.83.1057,kane3,PhysRevB.76.045302, RevModPhys.82.3045}. The electronic behavior of periodic materials is studied by the Bloch theory, that defines the bands below the Fermi level in the first Brillouin zone \cite{shen2012topological,asboth2016short,KANE20133}. The invariant is computed by the study of the so-called Berry–Pancharatnam–Zak phase, which is a geometric phase of the eigenstates associated with each band \cite{ghatak2019new,palumbo2019tensor,article4,blanco2020}. The clearest manifestation of the topological phase is the metallic states on the surface or edges, which are protected by time reversal symmetry. An important characteristic of these states is the linear dispersion relation, which makes them attractive for their application in high-speed electronics devices \cite{PhysRevLett.119.106602,article3, electronics7100225}. The helical propagation has spin-momentum locking property for application in heterostructures with magnetic materials in technologies such as spintronics \cite{ electronics7100225, Rachel_2018, CHONG2021100939, PhysRevB.95.205422}.\\

%Similar to the quantum hall effect, topological insulators (TI) present helical electronic states at their surface but without the use of strong external magnetic fields \cite{kane3,RevModPhys.82.3045, asboth2016short}.

In the topological insulators group we have three dimensional materials, whose studies are mainly focused on $Bi_{2}Se_{3}$ belong to the family of the strong spin-orbit coupling (SOC) TIs \cite{PhysRevB.95.205422, zhang2010first, manjon2013high}. Its structure is based on the so-called quintuple layers (QL) which makes them useful to produce samples with different thicknesses in a controlled manner, ideal for the synthesis of heterostructures \cite{Claro2021,ajeel2017topological}. In 2D materials, topological effects appear in layers with a strong spin orbit interaction. It is also explored in $HgTe/(Hg, Cd)Te$ quantum wells and $InAs/GaSb$ heterostructures \cite{kane3, 2016,ren2020engineering, graf2013bulk, gusev2011transport, culcer2010two}. 
Nanotubes, nanowires or low-dimensional heterostructures are one-dimensional examples. In the Su-Schrieffer-Heeger (SSH) tight-binding model for insulators, the Peierls distortion generates topologically protected solitonic states \cite{RevModPhys.83.1057,shen2012topological, asboth2016short, li2015winding}. Topological properties are explored from a two-level wavefunction as is mentioned in detail in \cite{simon2018}. This model with staggered Zeeman fields and electric potentials periodically modulated used to study of Floquet edge states, with applications in quantum computing as shown in \cite{L2019}. In superconducting systems, the  topological study predicts the appearance of isolated pairs of Majorana zero modes, on one-dimensional p-wave superconductor described by Kitaev chain model \cite{RevModPhys.83.1057, asboth2016short, qi2010chiral, xu2014artificial}. These states are of great interest for the development in quantum computing which projects them into future technological applications.  \cite{RevModPhys.83.1057, KANE20133, article4, article3, electronics7100225, qi2010chiral}.\\

 Different methods make it possible to build heterostructures with $Bi_{2}Se_{3}$ thin films. The number of QLs defines the electronic mobility at the edge due to the hybridization of the states of each edge \cite{lang2013competing,wang2011topological}. Magnetoresistance and conductance measurements show a drop in the surface current of thin layers of $Bi_{2}Se_{3}$  with less than 4 QLs, which correspond to an approximate thickness of $4 nm$  \cite{lu2016weak,brahlek2015transport,  park2010robustness}. In the case of a few QLs, hybridization between the states increases and this causes a gaped energy bands, that behave as a trivial insulator. In this frame, the $In_{2}Se_{3}$ is of great interest since it belongs to the same structural family of $Bi_{2}Se_{3}$ of QLs, but  measurements at the edge by ARPES and low energy models show a trivial insulator \cite{ li2018large, li2021low, collins2020electronic}, this makes them relevant materials to build heterostructures with combined topological properties \cite{ Rachel_2018, CHONG2021100939, PhysRevB.95.205422, hoffman1989negative, ballet2014mbe,  PhysRevA.98.043838,  PhysRevLett.108.220401}. For example, \cite{Shibayev2019} proposes a macroscopic global phase model where layers of $Bi_{2}Se_{3}$ and $In_{2}Se_{3}$ are combined to form a binary superlattice in configurations with a different number of QLs. They show topological phase diagrams verified experimentally by measurements of low-temperature magneto-transport for weak antilocalization effects \cite{lang2013competing, lu2016weak}. A similar idea is presented in \cite{Li2014} with superlattices of $SnTe/CaTe$ that are crystalline topological insulators with multiple valley Dirac states. \cite{hsieh2012topological,tanaka2012experimental, li2013single}. According to the hybridization between the surface states and the configuration of the superlattice, they make an approximation to the numerical calculation of topological phase diagrams and suggest the artificial synthesis of weak and strong topological insulators. However, it would be interesting to build a more detailed model that allows us to know if when alternating topological and non-topological materials, the heterostructure remains topological or not; what are the optimal magnitudes and properties of the materials to obtaining topological heterostructures?. Through the coupling of different chains it is possible to study the local effects of the edge states and the interaction between materials with different topological properties, in addition to an analytical and numerical study of the transport properties through the differential conductance. Motivated by this, we propose a method to study heterostructures based on SSH chains in which we analyze different configurations that combine topological and non-topological materials. In particular, we calculate topological phase diagrams as a function of coupling parameters in SSH chains and binary superlattices with chains of different sizes, as well as the detailed characteristic edge states and differential conductance (DC). For this purpose we find the Green Function (GF) of  this heterostructures in a maximally localized Wannier functions basis \cite{RevModPhys.84.1419} in order to obtain the local density of states (LDOS) that allows characterizing their edge states and study the transport properties.\\

The article is organized as follows. In section 2, we calculate the GF of an arbitrary chain in the SSH model, as well as the GF and edge states of a heterostructure of chains coupled with different  topological properties, in which we study the LDOS. In section 3, we calculate the topological phase diagrams from the eigenstates of the bulk Hamiltonian, and we show an analytical relationship of the coupling parameters that allows finding the boundaries of the phase diagram and comparing them with those obtained in the previous section. Finally, in section 4 we discuss the transport properties of the edge states  using the non-equilibrium Green's function formalism in order to find the DC where we will analyze the tunneling of the edge states through the heterostructure.

%Starting from the an spinless Hamiltonian in a localized base, the Green function of the primitive cell is calculated and with the Dyson method the function of the edge of a chain of arbitrary size is obtained. The same method allows modeling the edge states structures based on chains with alternating topological and trivial properties. From the reading of the edge states, maps of the intensity of the peak are obtained for the different parametric configurations. Next we consider the topological phase diagrams obtained from the zak phase calculation for a superlattice written in a multiband Hamiltonian [??]. With the zero energy states where band inversion occurs, an analytical parametric relationship is also proposed that allows us to see the phase transition of the proposed structures, where the differential conductance of the structures is calculated, finding a high transmission of the edge states in the tunnel boundary.

\vspace{0.5cm}

\section{SSH model and Green functions }

The SSH is a one-dimensional nearest-neighbor tight-binding model with two sublattice sites represented by atoms A and B \cite{RevModPhys.83.1057, asboth2016schrieffer, xie2019topological, zhang2021topological,  meier2016observation}. The Hamiltonian in a base of spinless localized  orbitals \cite{RevModPhys.84.1419} of a chain of $N$ cells is given by:

\begin{equation}
H=\sum\limits^{N}_{i}(\varepsilon _{A}c_{Ai}^{\dagger}c_{Ai}+\varepsilon_{B}c_{Bi}^{\dagger}c_{Bi}+vc_{Bi}^{\dagger}c_{Ai}+wc_{Bi}^{\dagger}c_{Ai+1}+h.c.),
\end{equation}

\hspace{-0.45cm} where $\varepsilon_{A}$ and $\varepsilon_{B}$ are the self-energies of atoms $A$ and $B$, respectively. The hopping $v$ couples the atoms in the cell, while $w$ couples atoms between neighboring cells.   

\bigskip

\begin{figure}[ht]
\begin{center}
\advance\leftskip-3cm
\advance\rightskip-3cm
\includegraphics[keepaspectratio=true,scale=0.35]{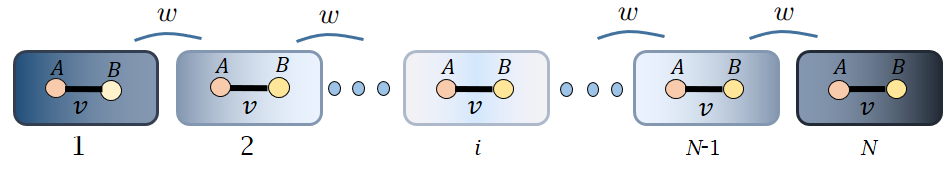}
\caption{Schematic representation of the SSH model. The circles denote A and B sublattice atoms with a hopping $v$. Each molecule is coupled with its neighbor cell by a coupling $w$.}
\label{gu3}
\end{center}\end{figure}

The topological properties of the chain are controlled by the intensity of the couplings $v$ and $w$. When $w<v$, we have a trivial insulating chain, while for $w>v$, the chain is topological. Figure \ref{gu3} shows a schematic representation of a one-dimensional chain of $N$ molecules. In the SSH model, all the  cell couplings are $v$ and all the  intercell couplings $w$ are the same \cite{asboth2016schrieffer}. However, in a more general version, the chain could have different values of hopping of each cell $v_{i}$ and $w_{i}$. With different configurations, we can study more complex models such as superlattice off-diagonal Harper model \cite{lado2019topological}. In order to explore the properties of a chain of finite length, we calculate the GF associated to  an isolated molecule with $A$ and $B$ sublattice atoms. The GF of this system is given by:

\begin{equation}
\hat{g}_{m}(E )=\frac {1} {(E - \varepsilon_A) (E - \varepsilon_B) - v^{2}}\left( 
\begin{array}{cc}
E -\varepsilon _{B} & v \\ 
v & E -\varepsilon _{A}%
\end{array}%
\right). 
\end{equation}

\bigskip

For simplicity, from now on we take self-energy values equal to zero ($\varepsilon_{A} = \varepsilon_{B}=0$), since in this model it acts as a gap.  In order to couple two of these molecules denoted by $\hat{g}_{11}$ and $\hat{g}_{22}$,  we use the Dyson equation approach through a self energy of the form: 

\begin{equation}
\hat{\Sigma}_{12}=w\left( 
\begin{array}{cc}
0 & 0 \\ 
1 & 0 %
\end{array}%
\right)=w\hat{C} ,
\end{equation}

\bigskip

\hspace{-0.5cm} where $\hat{\Sigma}_{12}=\hat{\Sigma}^{T}_{21}$. By solving the Dyson equation $\hat{G}=\hat{g}+\hat{g}\hat{\Sigma} \hat{G}$, we obtain the perturbed GF on the left edge given by

\begin{equation}
\hat{G}_{11}(E)=\hat{\Delta} (\hat{g}_{11}(E),\hat{g}_{22}(E)),
\end{equation}

\hspace{-0.45cm} where

\begin{equation}
\hat{\Delta} (\hat{a},\hat{b}_{})=(\hat{1}-\hat{a}\hat{\Sigma}
\hat{b}\hat{\Sigma}
)^{-1}\hat{a}.
\end{equation}

\vspace{0.5cm}

\hspace{-0.5cm} $\hat{\Delta}$ is an operator that depends on the GF of each structure to be coupled. In this case, we find the perturbed GF of two coupled molecules, so $\hat{g}_{11}=\hat{g}_{22}= \hat{g}_{m}$. Iteratively, we can compute the GF on the edge of a chain of arbitrary length.\\

\begin{figure}[ht]
\begin{center}
\advance\leftskip-3cm
\advance\rightskip-3cm
\includegraphics[keepaspectratio=true,scale=0.7]{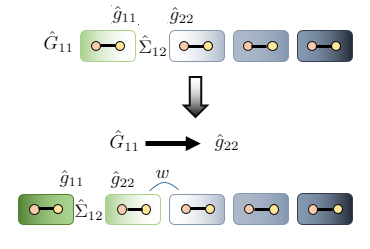}
\caption{Schematic representation of the method that allows us to find the GF at the edge of a chain of arbitrary length. Each molecule is coupled on the left side of the chain with self-energy $\hat{\Sigma}_{12}$. The perturbed GF $\hat{G}_{11}$ depends on $\hat{g}_{11}=\hat{g}_{m}$ and $\hat{g}_{22}$, which is the left edge of the chain.}
\label{ga16}
\end{center}\end{figure}

The schematic representation in Fig. \ref{ga16} shows the method  for coupling another cell to the chain. We initially calculate the perturbed GF $\hat{G}_{11}$ of coupling a molecule denoted by $\hat{g}_{11}$ to the left side of a chain with $\hat{g}_{22}$. When coupling a new molecule to the chain, the unperturbed GF replaces it with the one calculated in the previous perturbation 
$\hat{g}_{22}=\hat{G}_{11}$, while $\hat{g}_{11}$ is always the GF of one molecule $\hat{g}_{m}$. This is how we generate a system of arbitrary size through the GF on the edge:

\begin{equation}
\hat{G}_{11}(E )=\hat{\Delta} (\hat{g}_{m},\hat{g}_{22}),
\end{equation}

\hspace{-0.5cm} where $\hat{g}_{22}$ is the GF of the left edge of a chain of $N-1$ cells. From this function, we derive the LDOS given by:

\begin{equation}
    \rho_{11}(E)=\frac{-1}{\pi}\operatorname{Im}[Tr(\hat{G}_{11}(E+i\eta ))],
\end{equation}

 \hspace{-0.5cm} with $i\eta$ as  a positive imaginary part in the energy.

\begin{figure}[ht]
\begin{center}
\advance\leftskip-3cm
\advance\rightskip-3cm
\includegraphics[keepaspectratio=true,scale=0.4]{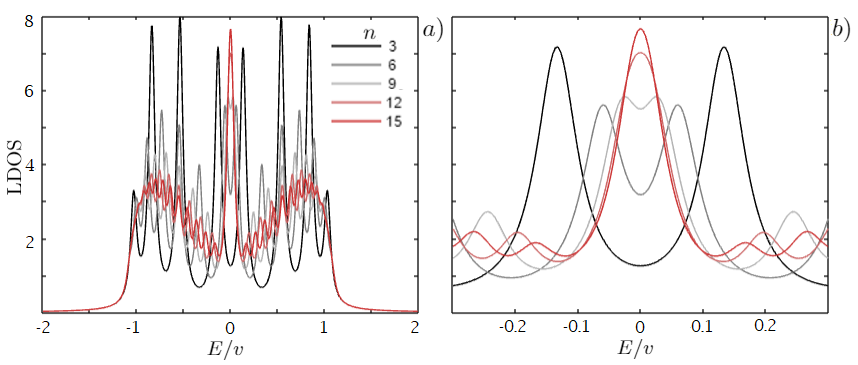}
\caption{(a) LDOS at the left edge of a finite chain with different sizes. (b) The same density in a short energy range. Both graphs have the parameters $w/v=1.2$.}
\label{gt11}
\end{center}\end{figure}

\begin{figure}[ht]
\begin{center}
\advance\leftskip-3cm
\advance\rightskip-3cm
\includegraphics[keepaspectratio=true,scale=0.4]{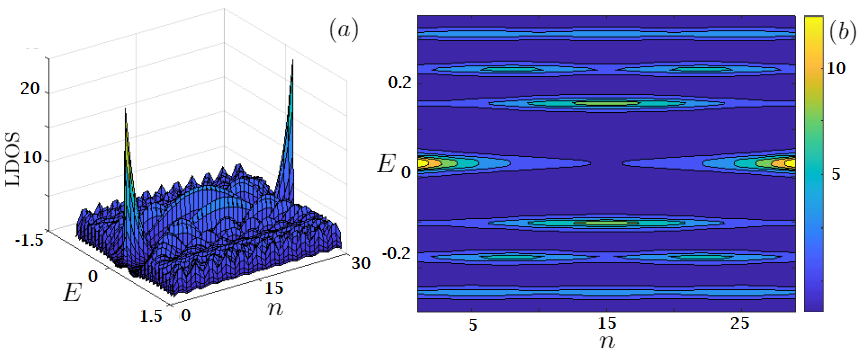}
\caption{(a) LDOS in function of the energy and the cell $n$ of a 30 molecules topological chain with $w/v=1.4$. (b) Contour lines.}
\label{gt9}
\end{center}\end{figure}

Figure \ref{gt11} shows the LDOS of a topological chain with $w/v=1.2$ and different numbers of molecules. In chains of a few cells, there is no zero energy peak because the states do not completely decay within the material. This generates a correlation between the states of each edge of the chain, and consequently the zero energy state disappears the LDOS. When the number of molecules increases, the correlation between both edges are negligible and the LDOS tends to a semi-infinite chain (see appendix A). To observe this phenomenon more clearly, we calculate the GF at the internal sites of the chain. We can also use Dyson's equation to obtain a LDOS inside the chain as a function of the cell $n$, as shown in Fig.\ref{gt9}.

On the Fig. \ref{gt9} (b), we observe that the zero energy states of a topological  chain of 30 cells are located at the edges, unlike nonzero energy states that are traveling modes. These states are associated with bulk band systems. The decrease in the intensity of the zero energy state inside the chain is associated with a localization length $ \lambda_{loc}$, which can be calculated directly from the LDOS  according to the value of the couplings $v$ and $w$, or from a low-energy effective model \cite{xie2019topological}. In this chain of 30 molecules, the intensity of the two states is zero after a few molecules in the chain, so they can be considered independent.\\

In order to study the hybridization between the edge states of each chain, it is useful to define the non-local GF $\hat{G}_{1N}(E)$, which gives the correlation between the states of the cell $1$ with those in $N$. From this function, we can calculate the probability of propagation of an electron between the edges of the chain and use it in section 4 to calculate transport properties in these systems. We use the same iterative method described in the previous section. According to the Dyson equation, the non-local GF  is given by
\bigskip

\begin{equation}
\hat{G}_{1N}(E)=\hat{\Omega} \prod_{i=1}^{N-1}(\hat{\Sigma}\Delta
(\hat{g}_{m}(E),\hat{g}_{ii}(E))), 
\label{gin}
\end{equation}

\hspace{-0.5cm} with

\begin{equation}
\hat{\Omega}= \hat{\Delta} (\hat{g}_{m}(E),\hat{g}_{m}(E))\hat{\Sigma} 
g_{m}(E).
\end{equation}

The operator $ \hat{\Omega}$  depends only on the GF of the molecule $\hat{g}_{m}(E)$ defined in the equation (2) and with $\hat{\Sigma}_{N-1,N}=\hat{\Sigma}_{12}=\hat{\Sigma}$.  The method to obtain the GF at the edge of a chain from the function of one molecule $g_{m}(E)$ has the advantage of preserving the matrix order by increasing chain size, unlike the Hamiltonian approximation that increases its matrix order according to the length of the chain. This feature also be useful for the calculation of the DC that we will describe in section 4.

\subsection{Heteroestructures}

\begin{figure}[ht]
\begin{center}
\advance\leftskip-3cm
\advance\rightskip-3cm
\includegraphics[keepaspectratio=true,scale=0.48]{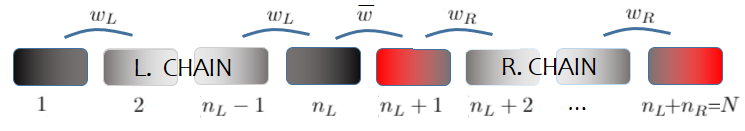}
\caption{Schematic representation of a two-chain coupled supercell. With the same value of $v$ for all molecules, the topological characteristics depend on the values $w_{L}$ and $w_{R}$ and on the sizes of the chains $n_{L}$ and $n_{R}$ (with $N=n_{L}+n_{R}$) , respectively, including a coupling potential between the chains $\bar{w}=\frac{1}{2}(w_{L}+w_{R})$}.
\label{gt12}
\end{center}\end{figure}

 By analogy to the construction of the local GF of a chain, we now do the calculations for a supercell of two coupled chains. The heterostructure can be larger if we perform the coupling of $M$ supercells. A schematic representation of the supercell is shown in Fig.\ref{gt12}.

\vspace{0.5cm}

The heterostructure consists of a supercell of $N$ molecules composed of two chains $L$ and $R$ of $n_{L}$ and $n_{R}$  cells coupled by hopping $w_{L}$ and $w_{R}$, respectively with $N=n_{L}+n_{R}$. The self-energy that couples the two chains at their edges depends on the parameter $\bar{w}$, which we take as the average of the $w_{L}$ and $w_{R}$. By applying the method explained in the previous section, we obtain the GF of the edge of the supercell as follows:

\begin{equation}
\hat{G}_{11}= \hat{g}_{11}+\hat{g}_{1,n_{L}}\hat{\Sigma} \hat{\Delta}(\hat{g}_{n_{L},n_{L}},\hat{g}_{n_{L}+1,n_{L}+1}) \hat{\Sigma} \hat{g}_{n_{L},1.} 
\label{g67}
\end{equation}

The unperturbed functions $\hat{g}_{11}$ and $\hat{g}_{n_{L},n_{L}}$ are the GF of each edge of the left chain. The operator $\hat{\Delta}(\hat{g}_{n_{L},n_{L}},\hat{g}_{n_{L}+1,n_{L}+1})$
performs the coupling (with self-energy $\bar{w}\hat{C}$) between this chain and the edge of the other chain with the function $\hat{g}_{n_{L}+1,n_{L}+1}$. The non-local GFs  $\hat{g}_{1,n_{L}}$ and $\hat{g}_{n_{L},1}$ appear in Eq. (10)  because the coupling is performed in a different cell from where we calculate GF. Note in Eq.(\ref{g67}) that in general terms $G_{11}$ connects the left chain with the rest of the heterostructure, thus the GF $\hat{g}_{n_{L}+1,n_{L}+1}$ can represent a single chain, or a pair of chains coupled in a previous process. This allows obtain the GF at the edge of a heterostructure with a number $M$ of supercells. In particular, we built a  periodic heterostructure of eight coupled chains ($M=4$).  The system is represented by a supercell composed of two chains, $L$ and $R$, as shown in Fig.\ref{gt12}. \\

\begin{figure}[ht]
\begin{center}
\advance\leftskip-3cm
\advance\rightskip-3cm
\includegraphics[keepaspectratio=true,scale=0.38]{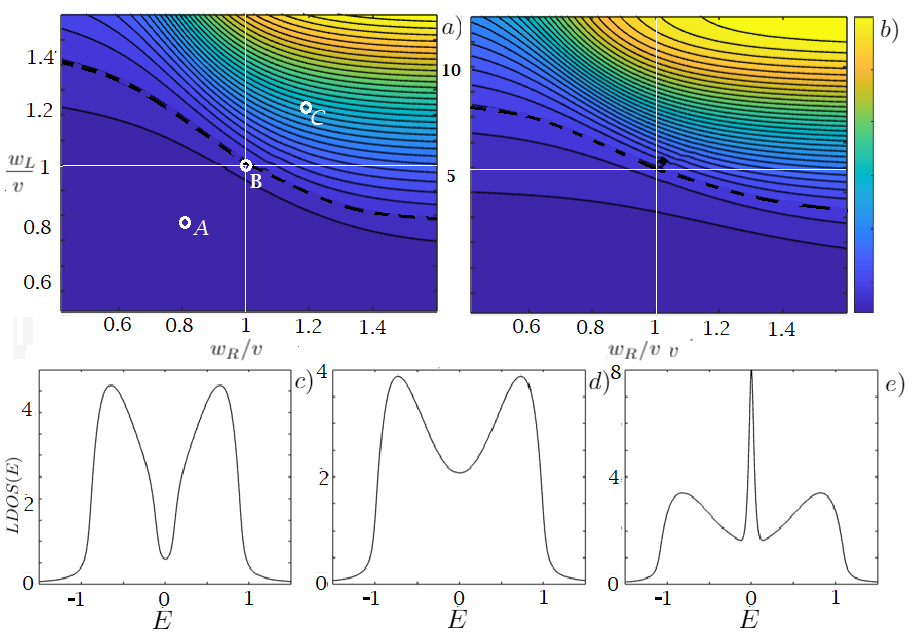}
\caption{Zero energy state mapping in LDOS that allow us to define topological and trivial regions similar to a phase diagram.(a) $n_{R}=3$ and $n_{L}=4$, (b) $n_{R}=4$ and $n_{L}=3$. (c) (d) and (e) are the LDOS at the edge for the configuration A, B and C, respectivelly.}
\label{gu1}
\end{center}\end{figure}

By analyzing the LDOS at the edge of the heterostructure evaluated at $E = 0$, we can generate intensity maps of this state as a function of the possible configurations of the parameters $w_{L}/v$ and $w_{R}/v$, as observed in Fig.\ref{gu1}. The shape of the map depends on the size of the chains. Fig.\ref{gu1} (a) shows the map of a binary heterostructure with $n_{L}=3$ and $n_{R}=4$. This map looks different from the system in (b) generated with $n_{L}=4$ and $n_{R}=3$. This occurs because the left chain, where we calculate the LDOS at the edge, has more cells and the edge state is better defined in it. To observe this, we separated the maps of the figure into quadrants. For example, the zone where $w_{L}/v>1$ and $w_{R}/v>1$  means both chains are topological. Conversely,  in the zone where $w_{L}/v<1$ and $w_{R}/v<1$, both chains are non-topological. In this zones, corresponding to quadrants I and III, respectively, it is trivial to determine the topological character of the structure. The LDOS shows a well-defined peak at $E = 0$ when both chains are topological, but when both chains are non-topological the peak does not appear. The interesting quadrants are II and IV, where topological and non-topological chains are present. In these parametric zones, we define as a criterion to separate the configurations with non-trivial topology with the value of the LDOS at $E=0$. In topological structures the zero energy state is a peak, as we can see in Fig.\ref{gu1}(e). In the maps, the configurations that are above the dotted curve correspond to topological ones, while those below are non-topological. We present this behavior as the first criterion to find the topological phase diagrams, taking into account the intensity of the edge state. In the next section, we find the topological phase diagram from the calculation of the topological invariant.

\section{Topological phase }

In this section, we find the topological invariant of heterostructures  from the calculation of the Berry–Pancharatnam–Zak phase. To achieve this, we impose boundary conditions on the supercell with $c_{N+1}=c_{1}$. This allow us to building phase diagrams by tuning parameters $w_{R/L}/v$ of two chains with sizes $n_{L/R}$ and comparing them with the maps calculated in the previous section. The Zak phase is defined from the occupied states in a discretized First Brillouin Zone as: 
\bigskip

\begin{equation}
    \theta=\sum_{i}^{m}\theta_{i}=\sum_{i}^{m}Im(\log (\prod_{l=1}^{J}\langle \langle
u_{i,k_{l}}|u_{i,k_{l+1}}\rangle\rangle)),
\end{equation}

\bigskip

\hspace{-0.65cm} where $m$ is the number of bands of the states with energy below the Fermi level $E_{F}$. The variable $k$ in the reciprocal space is discretized in $J$ parts. The condition of periodicity imposed on real space allows describing the system in the FBZ with $ k_ {J + 1} = k_ {1} $. The topological invariant takes 3 characteristics values: zero if the system has trivial topology and $\pm \pi$  for non-trivial heterostructures. Tuning in the coupling parameters $ w_ {L}/v$ and $w_ {R}/v$ generates a phase diagram that delimits the zones where the non-trivial phase is similar to the maps generated in the previous section.

\begin{figure}[ht]
\begin{center}
\advance\leftskip-3cm
\advance\rightskip-3cm
\includegraphics[keepaspectratio=true,scale=0.35]{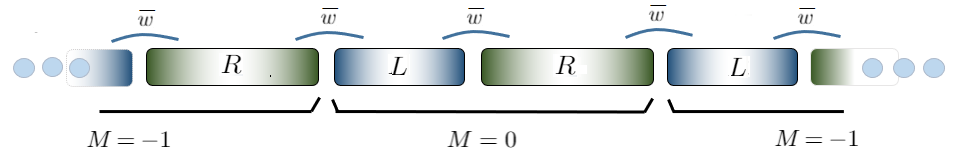}
\caption{Schematic representation of a superlattice made of two chains with different topological properties.}
\label{ga18}
\end{center}\end{figure}

 In Fig.\ref{ga18}, we show
 a schematic representation of the infinite superlattice defined by a supercell that is composed of two chains. The change of couplings $ w_ {L}/v$ and $w_ {R}/v$ gives each of the settings.

\subsection{Cross band parametric condition}

In this part, we present a  useful identity that allows finding the topological phase diagram. The  line that delimits the parametric zones where the heterostructure is topological, represents the phase transition due to the variation of the coupling parameters. In the intensity map of the zero energy state, a criterion was defined from the maximum of the LDOS close to $E=0$. Here, the eigenvalues of the Hamiltonian are used to find the parametric relationship that generates the characteristic cross bands. We perform the analysis in a reciprocal space where the Hamiltonian is defined in the following form:  
 
 \vspace{0.5cm}
 
$$ 
 H_{k}=\left( 
\begin{array}{ccccccc}
0 & v_{1} & 0 & 0 & \cdots & 0 & w_{N}e^{-ikL} \\ 
v_{1} & 0 & w_{1} & 0 &  &  & 0 \\ 
0 & w_{1} & 0 & v_{2} & &  & \vdots \\ 
0 & 0 & v_{2} & 0 & \ddots&  &  \\ 
\vdots &  & & \ddots &  \ddots &  & 0 \\ 
0 &  &  &  &  & 0 & v_{N} \\ 
w_{N}e^{ikL} & 0 & & \cdots & 0 & v_{N} & 0%
\end{array}%
\right) \\
$$

\vspace{1cm}

\hspace{-0.5cm} where $L=Na$ is the size of the supercell. This Hamiltonian has eigenvalues that represent the energy bands $ E_ {m} (k) $. Since we are interested in $E=0$ states, the wave equation of the system takes the form: 

\begin{equation}
    H |u_{i}\rangle=0.
\end{equation}

From this homogeneous equation we find a parametric relationship for the cross bands which, in general, is obtained from the determinant equal to zero $Det(H)=0$

\begin{equation}
   \prod_{i=1}^{N}v_{i}=\prod_{i=1}^{N}w_{i}.
\end{equation}

This result is general for any chain with couplings $v_{i}$ and $w_{i}$ and can be used in our supercell system composed of two coupled chains. Here, all the hoppings $v_{i}$  are equal ($v_{i}=v$), and the parameters $w_{i}$ are divided between those of the left and right chain $w_{L/R}$ according to the size of each chain. The parameter $\bar{w}$ that couples the chains is taken as the average between $w_{L}$ and $w_{R}$. Then,  when we have two chains of arbitrary lengths $ n_ {L} $ and $ n_ {R} $, the parametric relationship takes the form:

\begin{equation}
    w^{n_{L}-1}_{L}w^{n_{R}-1}_{R}\bar{w}^{2}=v^{N}.
\end{equation}

A particular solution of this equation is given by assuming that the parameter that couples both chains is constant $\bar{w}=v$. In this case we have

\begin{equation}
    \frac{w_{L}}{v}=(\frac{v}{w_{R}})^{\frac{n_{R}-1}{n_{L}-1}}. 
\end{equation}

\hspace{-0.5cm} This equation shows an inverse relationship in a graph of $w_{L}/v$ vs  $w_{R}/v$. In case the chains are of equal size, the curves are hyperbolas with asymptotes on the axes.\\

\begin{figure}[ht]
\begin{center}
\advance\leftskip-3cm
\advance\rightskip-3cm
\includegraphics[keepaspectratio=true,scale=0.6]{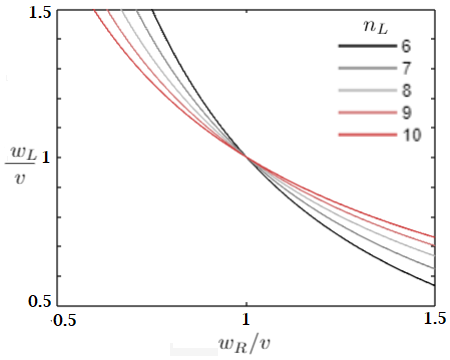}
\caption{Parametric relationship of the hopping $w_{R}$ and $w_{L}$ when hopping between chains is $\bar{w}=v$. The right chain has a fixed size of $n_{R}=8$ coupled with a left chain of different sizes.}
\label{gu2}
\end{center}\end{figure}

In case $w_{L}=w_{R}=w$, we obtain the band crossing relation for the SSH model in which $w = v$. Figure \ref{gu2} shows the curves generated by a right chain of 8 melecules coupled with a left chain of variable size. These curves delimit the zones where we have a topological and non-topological phase. For the configuration that interests us, and that we built in the previous section, the coupling between the chains is given by:

\begin{equation}
    \bar{w}=\frac{w_{L}+w_{R}}{2}. 
\end{equation}

\begin{figure}[ht]
\begin{center}
\advance\leftskip-3cm
\advance\rightskip-3cm
\includegraphics[keepaspectratio=true,scale=0.37]{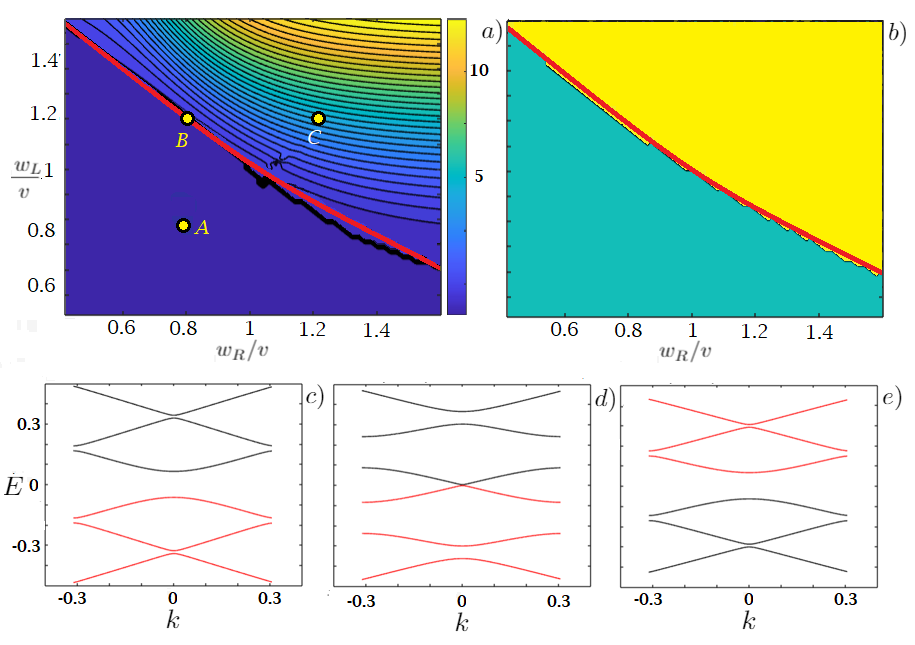}
\caption{(a) Phase diagrams of the edge state intensity in LDOS for $n_{L}=6$ and $n_{R}=4$. (b) Diagram of the topological invariant obtained through the Zak phase. The red line in both maps is obtained from the parametric relationship of band closure of Eq. (14). (c) (d) and (e) present the electronic spectrum of the superlattice at different points on the diagram. In these spectra, we have bands crossing in the edge region of the phase diagram.}
\label{gt1}
\end{center}\end{figure}

In this way, we find a functional relationship for the coupling parameters of identical systems to those built from the intensity of the $E=0$ state. Fig.\ref{gt1} presents the intensity map of the zero energy state in the LDOS, in contrast with the topological phase diagram obtained from the Zak phase. The curve of the parametric relationship was drawn in red on each graph.\\

Initially, both graphs exhibit similar zones. The red curve on the phase diagram perfectly overlaps the boundary denoted by the diagram. This is an expected result, since both approaches come from the same Hamiltonian. The same curve placed on the edge state intensity map does not delimit the zones with the same approximation as in the phase diagram. However, it is very close to the criterion used to delimit the topological zones of quadrants II and IV, as discussed in the cases of Fig.\ref{gu1}.\\

We can also observe this from the spectral relationship calculated in different areas of the map. In quadrants I and III, we obtain similar bands to structures based solely on topological or non-topological chains,  with an open gap that characterizes the bulk states. When we compute the energy spectrum at a point on the red transition line, the cross bands, which characterizes the transition. Consequently, we can conclude the agreement in the bulk surface relationship. It also validates the criteria used in the previous section to approximate the topological phase diagram from reading the intensity of the zero energy states in LDOS.

\section{Electric transport }

In this section, we focus on the analysis of the differential conductance of heterostructures that were described in the previous sections. We couple two electrodes at left and right edges of the heterostructure. As in any experimental setup in which electrical properties are measured, the structure is connected to electrodes of much larger dimensions. For this purpose, we model the GF of the edge of each electrode as monatomic semi infinite chains ($v=w\equiv t$) (see appendix A). Between the electrodes, there is a potential difference $V$ generating a current of charge carriers through the heterostructure. We also calculate the local conductance from the left electrode to the heterostructure.  The transport properties are calculated from the formalism of non-equilibrium Green functions, and the local and non-local GF already defined in Eq.(6) and Eq.(8). The DC is calculated from the current function, which is given by:

\begin{equation}
   I=\frac{2e}{\hbar}\int dE T(E,eV)(n_{F}(E-eV)-n_{F}(E)),
\end{equation}

\hspace{-0.5cm} where $n_{F}(E)$ is the Fermi-Dirac distribution, $V$ is the potential difference between the electrodes and $e$ is the charge of the electron. At zero temperature, the conductance is proportional to the transmission function given by:

\begin{equation}
    \sigma=\frac{2e}{\hbar}\frac{dI}{dV}=\frac{2e}{\hbar}T(eV),
\end{equation}

we obtain the local conductance $\sigma_{l}$ and nonlocal $\sigma_{nl}$:

\begin{equation}
    \sigma_{l}=\frac{8\pi^{2}e}{\hbar}t^{2}_{L}t^{2}_{R}\rho_{L}(eV)\rho_{11}(eV)|\hat{G}_{11}(eV)|^{2}.
\end{equation}

and

\begin{equation}
    \sigma_{nl}=\frac{8\pi^{2}e}{\hbar}t^{2}_{L}t^{2}_{R}\rho_{L}(eV)\rho_{R}(0)|\hat{G}_{1N}(eV)|^{2},
\end{equation}

\hspace{-0.5cm} Here, $\rho_{L/R}(E)=-\frac{1}{\pi}$Im(Tr($\hat{G}_{LL/RR}(E+i\eta)$)) corresponds to the LDOS of each electrode, and the non-local GF $\hat{G}_{1N}(eV)$ is perturbed by electrodes with $\Sigma_{L,1}=t_{L}\hat{C}$ and $\Sigma_{N,R}=t_{R}\hat{C}$. An analytic expression of this function is obtained with the Dyson equation in a similar way to the previous sections, and is given by:

 \begin{equation}
 \hat{G}_{1N}
  =\hat{\Lambda}_{L}\hat{g}_{1N}(\hat{1}-\hat{\Lambda}_{R}\hat{\Gamma}_{R}\hat{g}_{N1}\hat{\Lambda}_{L} \hat{g}_{1N})^{-1}\hat{\Lambda}_{R},
 \end{equation}
 
\hspace{-0.65cm} and 

\begin{equation}
\hat{G}_{11}=(\hat{g}_{11}^{-1}-\hat{\Gamma}_{L})^{-1},
 \end{equation}
 
\hspace{-0.65cm} with 

\begin{equation}
\hat{\Lambda}_{L}=(\hat{1}-\hat{g}_{11}\hat{\Gamma}_{L})^{-1}\ \ \text{and}\ \hat{\Lambda}_{R}=(\hat{1}-\hat{\Gamma}_{R}\hat{g}_{NN})^{-1}.
\end{equation}

\begin{figure}[ht]
\begin{center}
\advance\leftskip-3cm
\advance\rightskip-3cm
\includegraphics[keepaspectratio=true,scale=0.6]{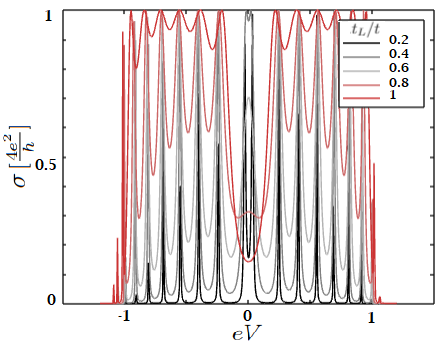}
\caption{ Non-local conductance of a topological chain of 10 molecules with $w/v=1.2$. The different curves represent the hopping with the electrodes due to value $t_{L}=t_{R}$.}
\label{gh8}
\end{center}\end{figure}

\hspace{-0.5cm} Here $\hat{\Gamma}_{L} =\hat{\Sigma}_{1L} \hat{g}_{LL}\hat{\Sigma} _{L1}$ and $\hat{\Gamma}_{R}=\hat{\Sigma} _{nR}\hat{g}_{RR}\hat{\Sigma}_{Rn}$  are the coupling terms between the structure and the electrodes. The non-local unperturbed GFs $\hat{g}_{1N}$ and $\hat{g}_{11}$ are the ones calculated in section 2. In general, the non-local GF can correspond to a single chain, to a heterostructure or a binary superlattice. The Fig.\ref{gh8} presents the electronic transmission of a single topological chain of 10 molecules for different values of the coupling level with electrodes $t_{L}=t_{R}$. \\ 

When the coupling of the structure with the electrodes is weak ($t_{L}/t=0.2$), we are at the tunnel limit  (black curve). This curve describes a transmission with a number of resonances equal to $N$. The resonances close to $eV = 0$ correspond to a topological edge state given by the parametric relation $w/v>1$. As shown in Fig.\ref{gh8}, this peak loses intensity when the value of the coupling between the heterostructure and the electrodes increases. In the transparent limit with a perfect coupling $(t_{L/R}/t=1)$, the intensity of this peak tends to zero (red curve), unlike resonances at $eV\not= 0$, which broaden. This indicates that border states have a distinctive characteristic and their transport occurs only by tunneling.\\

\begin{figure}[ht]
\begin{center}
\advance\leftskip-3cm
\advance\rightskip-3cm
\includegraphics[keepaspectratio=true,scale=0.4]{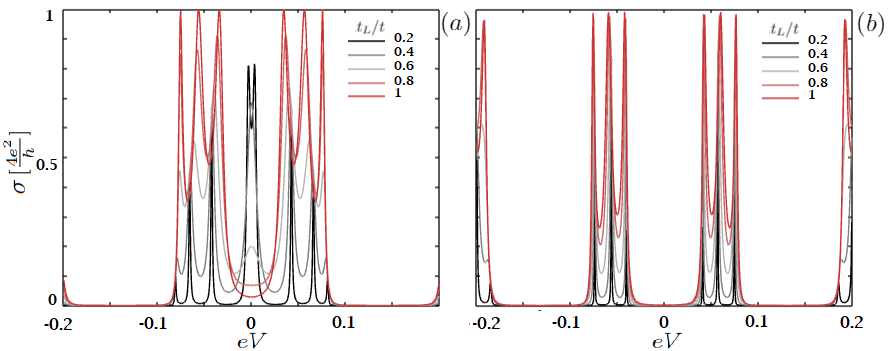}
\caption{Conductance of a heterostructure based on two chains of size $n_{L}=6$, $n_{R}=4$. (a) The largest chain is topological with $w_{L}/v=1.4$, while the shortest one is $w_{R}/v=0.8$. (b)  The largest chain is the non-topological with $w_{L}/v=0.8$, while the shortest one is topological with $w_{R}/v=1.4$. The different curves arise from varying the coupling $t_{L}=t_{R}$ until the transparent limit. }
\label{gt6}
\end{center}\end{figure}

In heterostructures, the formation of the peak at $E=0$ depends on the intensity of the couplings of each chain and the size of each one, as mentioned in the previous sections. In Fig.\ref{gt6}, we show the electronic transmission curve of two chains with $n_{L}$ and $n_{R}$ sizes, where each curve comes from the change in the parameters $t_{L}=t_{R}$.\\

\begin{figure}[ht]
\begin{center}
\advance\leftskip-3cm
\advance\rightskip-3cm
\includegraphics[keepaspectratio=true,scale=0.4]{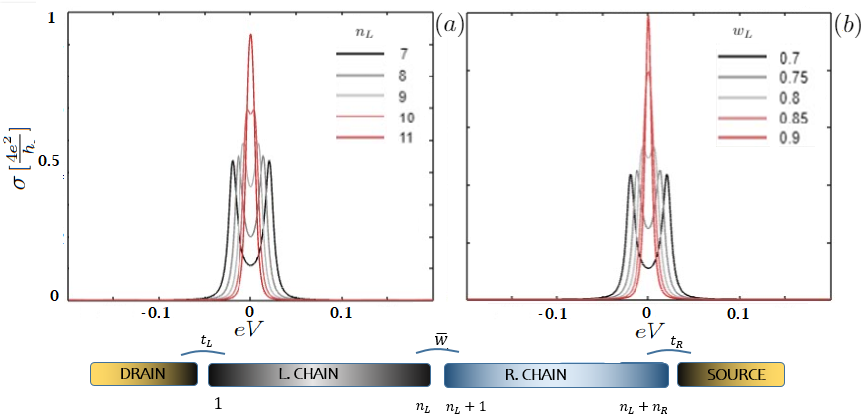}
\caption{Nonlocal conductance in the tunnel limit. (a) change in the length of a left topological chain, with a non-topological chain of a fixed size ($n_{R}=4$) from the right chain. (b) Two chains with fixed length ($n_{L}=4$, $n_{R}=6$) tuning the coupling parameter $w_{L}/v$ with a fixed $w_{R}/v=0.8$. In both figures, the resonance is formed at $E=0$.}
\label{gt8}
\end{center}
\end{figure}

 In the tunnel limit, the conductance shows resonance peaks homogeneously distributed over the voltage with a number of resonances proportional to the number of molecules in the heterostructure. In this case the transmission peak at $E = 0$ comes from the propagation of the edge state of the topological chain between the edges that make contact with the electrodes. Here we also see that the transmission peak of the tunnel boundary disappears at the transparent boundary, and this only happens with resonance at $E=0$. As in the LDOS, the DC depends on the size of each chain and the intensity of the couplings that define its topological character.  Likewise, the size of the non-topological chain is important, since the edge state decay inside this chain, this is reflected in an exponential decay of the peak in DC when we increase the size of the non-topological chain. Fig.\ref{gt8} shows how the conductance of the topological state changes under different parametric conditions.\\

A resonance peak in the conductance of the topological state is formed by increasing the size of the topological chain, which is evidence of the hybridization of the states of each edge as shown in Fig.\ref{gt8}(a). If the chain have a sufficient size to define the state, and we continue to increase the size of the chain, the conductance, which is proportional to the function $G_{1N}$, decays since the correlation between the states at edge $L$ and edge $R$ decreases. In Fig.\ref{gt8}(b), the size of both chains is fixed, and  the value $w_{L}/v$ is increased. We observe that the states are located more on the edges, which causes low correlation between states, thus allowing the state to be well defined at the edge in chains of few molecules. By constructing a non-local conductance map at zero energy we can study how the size of the chains affects the conductance. \\

\begin{figure}[ht]
\begin{center}
\advance\leftskip-3cm
\advance\rightskip-3cm
\includegraphics[keepaspectratio=true,scale=0.85]{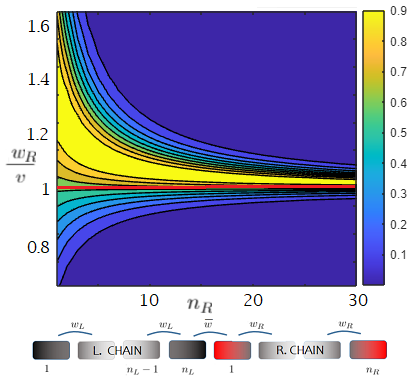}
\caption{DC at the tunnel limit of a fixed-size topological left chain ($n_{L}=8$ and $w_{L}/v=1.4$) coupled to a right chain with variable parameters. Below the red line the right chain is non-topological and the DC decays rapidly with increasing $n_{R}$. }
\label{gu12}
\end{center}\end{figure}

The map of Fig. \ref{gu12} shows the nonlocal conductance in the tunnel limit for $eV=0$, of two chains. The left chain is a fixed-size topology, with a well-defined peak on the edge. When the $L$ chain is connected to a non-topological chain (below the red line), the conductance peak reduces in intensity and decays with the size of the $R$ chain. Above the red line $(w_{R}/v>1)$, both chains are topological, the conductance maintains a high value even for a very large right chain. In both cases we see that the nonlocal DC tends asymptotically to zero, this is expected since the conductance depends on the distance between both contacts. Thus we find the optimal heterostructure configurations that allow us to obtain a high differential conductance\\

The local conductance is not affected by the size of the chains. We can see this in a conductance map of two coupled fixed-size chains, depending on the coupling parameters, similar to the one obtained in section 2 for the LDOS presented in Fig.\ref{gu1}. In the tunnel limit the zones of non-zero conductance  defining a map similar to the phase diagrams.

\begin{figure}[ht]
\begin{center}
\advance\leftskip-3cm
\advance\rightskip-3cm
\includegraphics[keepaspectratio=true,scale=0.37]{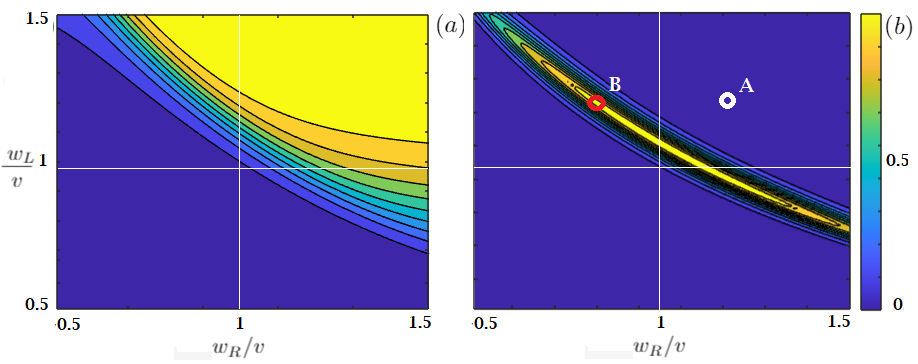}
\caption{(a) Intensity map of the local conductance of the zero energy state,  for different configurations of a heterostructure based on two chains of $n_{L}=6$ and $n_{R}=4$ molecules repeated $M=4$ times alternately. The map is similar to the one obtained for the LDOS in Fig \ref{gt1} . (b) non-local conductance for the same configurations. With the parameters of points A and B, the LDOS is calculated close to $E=0$ as shown in figure 15}
\label{gu8}
\end{center}\end{figure}

 Figure \ref{gu8} shows the conductance map in the case of a heterostructure with 4 coupled supercells defined in Fig.\ref{gu12} with the GF in the Eq.(6) to Eq.(8). In Fig.\ref{gu8}(b) the zone of high intensity in the conductance is drastically reduced. In most of the parametric configurations, the value of DC in the tunnel limit is zero according to Eq.(21). This is mainly due to the dimensions of the heterostructure, since for 4 coupled supercells, we have 40 molecules and this affects the correlation between their edges. 
This does not happen in the local conductance as we see in (a), since only the transmission through the edge is calculated and not over the entire heterostructure. In this case the the local conductance is proportional to LDOS of the heterostructure and its map is similar to the one found in the Fig.\ref{gt1} where we calculated the Zak phase. A map of LDOS as a function of energy and internal site $n$ of the chain, similar to Fig.\ref{gt9}, shows how the coupling of the different states at the edges causes a non-zero conductance.In the area corresponding to two non-topological chains, the transmission is zero since an edge state is not defined in any of the chains. The non-zero transmission configurations delimit similar zones to those found in the LDOS in Fig.\ref{gt1}.\\

Finally, we must mention the difference in the intensity of the DC in Fig.\ref{gu1} b). The line that separates the points with transmission different from zero is similar to the one that appears in the study of the topological properties of the previous sections in the Fig \ref{gt1} (a) and (b). From this curve, the intensity of the transmission begins to increase as the value of the coupling parameters changes. The different levels describe intensity variation curves similar to the initial curve. After reaching a maximum value, the transmission starts to decrease in high values of $w_{L}/v$ and $w_{R}/v$. In the first quadrant, we have a supercell of two topological chains and for high values of the coupling parameters, the transmission intensity is low because the state has a high intensity but decays faster inside the chain. Furthermore, the states appear only at the edges where there is a phase transition. Since all chains are topological, then, peaks only form at the edges of the heterostructure. This is a similar case to calculating a single chain with the size of the heterostructure. Being isolated enough, the correlation between the states depends on the coupling parameters configuration, hence conductance is reduced. This result is more evident when we couple several supercells, thus obtaining a larger heterostructure.\\

\begin{figure}[ht]
\begin{center}
\advance\leftskip-3cm
\advance\rightskip-3cm
\includegraphics[keepaspectratio=true,scale=0.47]{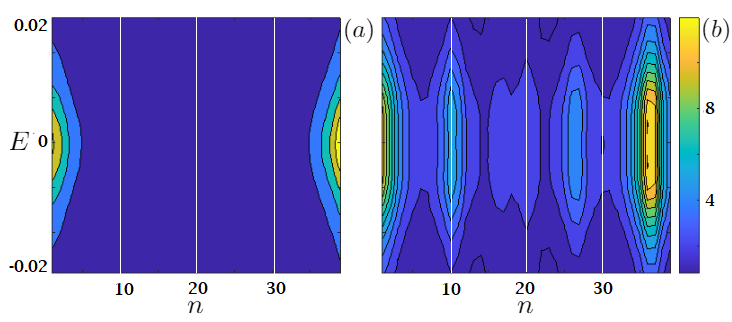}
\caption{LDOS as a function of the energy and the molecule $n$ of  a heterostructure with 4 coupled supercells. (a) A supercell of two topological chains $w_{L}/v=w_{R}/v=1.2$ (b) A topological chain with $w_{L}/v=1.2$ and a non-topological one $w_{R}/v=0.8$. The configurations correspond to point A and B of Fig. \ref{gu8}, respectively }
\label{gu10}
\end{center}\end{figure}

Point A in Fig.\ref{gu8} indicates that the DC is zero in a 4 coupled supercell heterostructure of two topological chains. The LDOS in a region close to $E=0$ for each site $n$ shows that the edge states are formed only at the edges as shown in Fig.\ref{gu10}(a). The DC tends to zero because the edge state decays rapidly inside the chain, unlike the LDOS with the configuration of point B in Fig.\ref{gu8}, which shows intensity peaks inside each supercell. This justifies its high value in DC, since it creates a collective effect that allows edge states to tunnel through the heterostructure.

\section{Conclusions}

We studied the topological and transport properties of chain-based binary heterostructures of the SSH model. We verified the parametric correspondence between the topological phase of an infinite system, and the intensity of the edge state obtained from the LDOS. From the Green function formalism and the Dyson equation, we  calculated the LDOS at the edge of the heterostructure. In different configurations, we found a peak at zero energy that represents a topologically protected metallic state. The intensity of this peak depends on the length of the chains $n_{L}$ and $n_{R}$, which determines hybridization between the states of each edge in a topological chain. For chains with few molecules, the edge state is not defined. This also depends on the relationship between the couplings $w_{R}/v$ and $w_{L}/v$ in the supercell.  With the intensity peak at zero-energy state tuning the coupling between molecules, we built maps of the different parametric configurations. The LDOS map shows a curve that separates topological heterostructures from those that are not.\\

The topological invariant obtained from the Berry–Pancharatnam–Zak phase shows diagrams as a function of the parameters $w_{R}/v$ and $w_{L}/v$ with a curve that separates topological configurations, similar to those found in the maps of LDOS at $E=0$. This shows a good agreement with the finite system approximation. We derived a usefull analytical relationship between the parameters to generate bands crossing in the energy spectrum. The curves calculated by numerical methods are the same as those found in the other approximations made. As a result we obtained a parametric characterization of the topological properties of the heterostructure.\\

In the transport properties calculated from the non-equilibrium Green functions formalism, we show that the differential conductance of these states through the heterostructure occurs only by tunneling, which is a property of topological states.  The conductance map at $eV=0$ describes high transmission zones similar to those calculated with the edge states and the topological phase. However, in heterostructures of several coupled supercells, the configurations that give high conductance are reduced due to the increased dimensions of the system. For the case of heterostructures with several supercells, the DC at $E=0$ can be used to find the topological phase transition since near to the transition line the edge states decay slowly and the conductance show a peak at $eV=0$.  We expect that our analysis can guide future experiments for the study of heterostructures based on alternanting different topological materials.
\\

\textbf{Acknowledgement}\\

We thank Dr. Pablo Burset for his comments and contributions that greatly improved the manuscript. Likewise appreciate the help of Dr. Shirley Gomez in support from the research project (QUIPU code: 202010027542), from this work derives.

\vspace{1.5cm}

\textbf{Appendix A: Green function for semi-infinite chain}

\vspace{1cm}

In this appendix, we show how to calculate the GF of a finite chain with the Dyson equation. We also derive an analytic expression for the GF of a semi-infinite chain. The process starts with molecule 1 being perturbed by the edge of a molecule or a chain of some size with unperturbed GF $g_{22}$ and coupling $\Sigma_{12}$, then, we obtained:\\

$\hat{G}_{11}=\hat{g}_{11}+\hat{g}_{11}\hat{\Sigma} _{12}\hat{G}_{21}$.\\

\hspace{-0.5cm} The perturbed GF $\hat{G}_{11}$ depends on the non-local GF $\hat{G}_{21}$ that should be calculated similarly. By applying Dyson equation again and knowing that unperturbed GF $\hat{g}_{21}$ is zero since before the coupling both molecules were separated, the following expression is achieved\\

\vspace{0.5cm}
$\hat{G}_{21}=\hat{g}_{22}\hat{\Sigma} _{21}\hat{G}_{11},$

\vspace{1cm}

\hspace{-0.5cm} then replacing:

\begin{equation}
    \hat{G}_{11}=(\hat{1}-\hat{g}_{11}\hat{\Sigma} _{12}\hat{g}_{22}\hat{\Sigma} _{21})^{-1}\hat{g}_{11}=\hat{\Delta} (\hat{g}_{m},\hat{g}_{22}).
\end{equation}

\vspace{0.5cm}

\hspace{-0.5cm} This equation can be solved analytically in a semi-infinite chain where $\hat{G}_{11}=\hat{g}_{22}$. In this case, we obtain a Green function of the form:

\vspace{0.5cm}

\begin{equation}
G_{11}(E)=\frac{D}{w}\left( 
\begin{array}{cc}
\frac{w}{AD}\left[ 1+\frac{v}{w}D\right] & 1 \\ 
1 & \frac{A}{v}%
\end{array}%
\right),
\end{equation}

\vspace{0.5cm}

\hspace{-0.5cm} with $A/B=(E - \varepsilon_{A/B})$. $D$ is an armonical function of energy and the parameters $v$ and $w$ 

\begin{equation}
 D=\frac{AB-(v^{2}+w^{2})}{2vw}-i\sqrt{1-\left[ \frac{%
AB-(v^{2}+w^{2})}{2vw}\right] ^{2}}.  
\end{equation}

\vspace{0.5cm}

The LDOS of the semi-infinite chain can be observed in Fig.\ref{gt11}. Now, we calculate the GF of a finite chain. For this we start from Eq. (22). As mentioned in section 2, the found function becomes unperturbed for the next coupling. For example, in a chain of 3 molecules, $g_{22}$ is the GF of a chain of two molecules, thus $g_{22}=G_{11}$ as shown in Fig. 2. Recursively, we can obtain GF on the edge of this chain:

\vspace{0.5cm}
$\hat{G}_{11}(E)=\hat{\Delta} (\hat{g}_{m}(E),\hat{g}_{22}(E)),$

\vspace{0.5cm}
\hspace{-0.5cm} with $\hat{g}_{22}(E)$ showing the same functional form, namely, 

\vspace{0.5cm}

$\hat{g}_{22}=\hat{\Delta} (\hat{g}_{m}(E),\hat{g}_{33}(E))$

\vspace{0.5cm}

\hspace{-0.5cm} therefore,

\vspace{0.5cm}
$\hat{G}_{11}(E)=\hat{\Delta} (\hat{g}_{m}(E),\hat{\Delta} (\hat{g}_{m}(E),\hat{g}_{33}(E)))$.

\vspace{0.5cm}

This procedure can be extended for a chain of size $N$. For chains of more than 4 or 5 molecules, the analytic expression is complicated, therefore, it was solved by numerical methods.

\vspace{1cm}

In order to find the non-local GF, we must do a recursive calculation. The perturbed GF $\hat{G}_{1N}$ arises from coupling cell $N$ with a chain of $N-1$ molecules. In a small system for $N=3$, for example, we have:

\vspace{1cm}

$\hat{G}_{13}=\hat{g}_{12}\hat{\Sigma} _{23}\hat{G}_{33},$

\bigskip

$\hat{G}_{33}=\hat{g}_{33}+\hat{g}_{33}\hat{\Sigma} _{32}\hat{G}_{23},$

\bigskip

$\hat{G}_{23}=\hat{g}_{22}\hat{\Sigma} _{23}\hat{G}_{33}.$

\vspace{1cm}

\hspace{-0.5cm} From here, we have:

\vspace{0.5cm}

$\hat{G}_{33}=(\hat{1}-\hat{g}_{33}\hat{\Sigma}_{32}\hat{g}_{22}\hat{\Sigma}_{23})^{-1}\hat{g}_{33},$

\bigskip

\hspace{-0.5cm} By substituting, we get the non-local GF:

\begin{equation}
\hat{G}_{13}=\hat{g}_{12}\hat{\Sigma}_{23}(1-\hat{g}_{33}\hat{\Sigma} _{32}\hat{g}_{22}\hat{\Sigma}_{23})^{-1}\hat{g}_{33}, 
\end{equation}

\vspace{0.5cm}

\hspace{-0.5cm} We still need to calculate functions $\hat{g}_{12}$ and $\hat{g}_{22}$, with a similar functional form:

\begin{equation}
    \hat{G}_{12}=\hat{g}_{11}\hat{\Sigma} _{12}(\hat{1}-\hat{g}_{22}\hat{\Sigma} _{21}\hat{g}_{11}\hat{\Sigma}
_{12})^{-1}\hat{g}_{22}.
\end{equation}

\bigskip

\hspace{-0.5cm} More generally, for a chain of $N$ molecules, we can write the GF as:

\begin{equation}
\hat{G}_{1N}=\hat{g}_{1,N-1}\hat{\Sigma}(\hat{1}-\hat{g}_{NN}\hat{\Sigma}\hat{g}_{N-1,N-1}\hat{\Sigma})^{-1}\hat{g}_{NN},
\end{equation}

with

\begin{equation}
 \hat{G}_{1N}=\hat{g}_{1,N-1}\Delta(\hat{G}_{N-1,N-1}),   
\end{equation}

\hspace{-0.5cm} and 

\begin{equation}
  \Delta(\hat{g}_{N-1,N-1})=\hat{\Sigma}(\hat{1}-\hat{g}_{NN}\hat{\Sigma}\hat{g}_{N-1,N-1}\hat{\Sigma})^{-1}\hat{g}_{NN}.  
\end{equation}

\vspace{0.5cm}

In this way each GF is computed until performing the calculation of a chain with two molecules:

\begin{equation}
\hat{g}_{1N}=\hat{\Omega}
(\hat{g}_{m})\prod_{i=1}^{N-1}\Lambda
(\hat{g}_{ii}^{r}), 
\end{equation}

\hspace{-0.5cm} with

\begin{equation}
\hat{\Omega}= \hat{\Delta} (\hat{g}_{m}(E),\hat{g}_{m}(E))\hat{\Sigma} 
g_{m}(E).
\end{equation}

\vspace{0.5cm}

\textbf{Appendix B: Non local Green function and transmission coefficient }

\vspace{1cm}

Here, we show how to calculate the GF of a structure perturbed by edge electrodes, which allows expressing the differential conductance from a transmission function. We perform the coupling of the non-local GF to the electrodes. We start with the left electrode, by applying the Dyson equation to perform coupling with autoenergy $\hat{\Sigma} _{1L}$: 

\vspace{0.5cm}

$\hat{G}_{1N}=\hat{g}_{1N}+\hat{g}_{11}\hat{\Sigma} _{1L}\hat{g}_{LL}\hat{\Sigma} _{L1}\hat{G}_{1N},$

\vspace{0.5cm}

\hspace{-0.5cm} where $\hat{g}_{LL}$ is the GF of a semi-infinite monatomic chain described in appendix A. Defining $A=\hat{\Sigma} _{1L}\hat{g}_{LL}\hat{\Sigma} _{L1}$,

\vspace{0.5cm}

\begin{equation}
 \hat{G}_{1N}=(\hat{1}-\hat{g}_{11}A)^{-1}\hat{g}_{1N}.   
\end{equation}\\

Now, for coupling with the right electrode we proceed in a similar way: 

\begin{equation}
   \hat{G}_{1N}=\hat{g}_{1N}+\hat{G}_{1N}B\hat{g}_{NN},
\end{equation}

\vspace{0.5cm}

\hspace{-0.5cm} By defining $B=\hat{\Sigma} _{NR}\hat{g}_{RR}\hat{\Sigma} _{RN}$, and $\hat{g}_{RR}$, the GF of the right electrode,

\begin{equation}
    \bigskip \hat{G}_{1N}=\hat{g}_{1N}(1-B\hat{g}_{NN})^{-1}.
\end{equation}

\hspace{-0.5 cm} It is worth mentioning that in Eq. 32 the unperturbed GF $\hat{g}_{1N}$ corresponds to the perturbed function $G_{1N}$ of Eq. 33 that already includes the left electrode. Finally we calculate $\hat{g}_{NN}$ which is equivalent to $\hat{G}_{NN}$ that couples the chain with the left electrode. The calculated GF now becomes unperturbed functions to perform the Dyson of the total coupling:

\vspace{0.5cm}

$\hat{G}_{NN}=\hat{g}_{NN}^{c}+\hat{g}_{N1}^{c}A\hat{g}_{1N}^{c}(\hat{1}-A\hat{g}_{11}^{c})^{-1}.$

\vspace{0.5cm}

\hspace{-0.5cm} Here, $g^{c}_{ij}$ are the functions of the structure when it has not been attached to any electrode calculated in appendix A. In summary,

\begin{eqnarray*}
\hat{G}_{1N} &=&\hat{g}_{1N}(1-B\hat{g}_{NN})^{-1} \\
\hat{g}_{1N} &=&(\hat{1}-\hat{g}_{11}^{c}A)^{-1}\hat{g}_{1N}^{c} \\
\hat{g}_{NN} &=&\hat{g}_{NN}^{c}+\hat{g}_{N1}^{c}A\hat{g}_{1N}^{c}(\hat{1}-A\hat{g}_{11}^{c})^{-1}.
\end{eqnarray*}

\vspace{0.5cm}

\hspace{-0.5cm} By Substituting the equations and applying matrix commuting properties we finally obtain:

\begin{equation}
    \hat{G}_{1N}=\left[ ((1-B\hat{g}_{NN}^{c})\left( \hat{g}_{1N}^{c}\right)
^{-1}(\hat{1}-\hat{g}_{11}^{c}A)-B\hat{g}_{N1}^{c}A\right] ^{-1}.
\end{equation}

With this function we calculate the DC described in Eq.(18). This conductance can be expressed in terms of the electronic transmission as shown in \cite{gomez, vasilesca, Paulsson2003ResistanceOA}.

\begin{equation}
    T(eV)=4\pi^{2}t^{2}_{L}t^{2}_{R}\rho_{L}(eV)\rho_{R}(0)|\hat{G}_{1N}(eV)|^{2}.
\end{equation}

%\begin{figure}[ht]
%\begin{center}
%\advance\leftskip-3cm
%\advance\rightskip-3cm
%\includegraphics[keepaspectratio=true,scale=0.4]{ga13.png}
%\caption{a) Transmission at the transparent limit of a topological chain of different leng.nl=11 wl=0.7 wr= 0.4 b) DOS for $ E = 0 $ in function of the internal site of the chain of 20 sites, }
%\label{visina8}
%\end{center}\end{figure}

%% Fundamentos de la fase topológica

\nocite{*}
%\stylecitereferences{natbib}

\medskip

\printbibliography

%\bibliography{aipsamp}

%\bibliography{cas-refs}

%\vskip3pt

\end{document}